\documentclass[10pt,a4paper,superscriptaddress,aps,prd,showkeys,showpacs,nofootinbib,reprint]{revtex4-1}
\usepackage[utf8]{inputenc}
\usepackage{verbatim}
\usepackage{amsmath,amsfonts,amssymb}
\usepackage[breaklinks,colorlinks]{hyperref}
\usepackage{natbib}
\usepackage{tikz}
\usepackage{float}
\usepackage{graphicx}
\usepackage{adjustbox}
\usepackage{tabularx}
\usepackage{amsbsy}
\hypersetup{%
,urlcolor=blue
,citecolor=blue
,linkcolor=blue
}

\begin{document}

\title{Optical properties of dynamical axion backgrounds}

\author{Jamie~I.~McDonald}
\email[]{jamie.mcdonald@tum.de}
\affiliation{%
 Physik-Department, James-Franck-Stra{\ss}e,  Technische Universit{\"a}t M{\"u}nchen,
	85748 Garching, Germany}

\author{Lu{\'i}s B.~Ventura}
\email[]{lbventura@ua.pt}
\affiliation{%
Departamento de F{\'i}sica da Universidade de Aveiro and CIDMA, Campus de Santiago, 3810-183 Aveiro, Portugal}%

\date{\today}

\begin{abstract}

We discuss spectral distortions,  time delays and refraction of light in an axion or axion-plasma background. This involves solving the full set of geodesic equations associated to the system of Hamiltonian optics, allowing us to self-consistently take into account the evolution of the momentum, frequency and position of photons. 
We support our arguments with analytic approximations and full numerical solutions.
Remarkably, the introduction of a plasma enhances the sensitivity to axion-induced birefringence,
allowing these effects to occur at linear order in the axion-photon coupling even when the axion background is not present at either the emission or detection points. 
This suggests a general enhancement of axion-induced birefringence when the background refractive index is different from one.
\end{abstract}

\maketitle

\section{Introduction}

Axions remain promising candidates for beyond the Standard Model physics. The QCD axion emerges as a solution to the strong-CP problem \cite{Peccei:1977hh,Weinberg:1977ma,Wilczek:1977pj} whilst string theory extensions of the standard model predicate a plethora of axion like particles (ALPs) \cite{Conlon:2006tq,Svrcek:2006yi}. Axions could also solve the dark matter problem  \cite{Preskill:1982cy,Marsh:2015xka,Hui:2016ltb}. The masses of axions can span a large range of scales, with Compton wavelengths on the order of galaxies down to table-top sizes of meters or centimeters. As a result, the axion is the subject of many current and proposed laboratory searches \cite{Asztalos:2009yp,TheMADMAXWorkingGroup:2016hpc,Millar:2016cjp,Majorovits:2016yvk,Anastassopoulos:2017ftl,Irastorza:2011gs,Dobrich:2013mja,Redondo:2010dp,Adler:2008gk}.

The aim of the present work is to explore the optical properties of an axion background due to its coupling to photons 
\begin{equation}\label{aGamGam}
\mathcal{L}_{a \gamma \gamma} = - \frac{g_{a\gamma\gamma}}{4} a F_{\mu\nu}\tilde{F}^{\mu\nu}\,, 
\end{equation}
where $g_{a\gamma\gamma}$ is the axion-photon coupling, $a$ is the axion field and $F_{\mu \nu}$ and $\tilde{F}_{\mu \nu}$ are the photon field strength and its dual, respectively. This interaction leads to non-trivial dispersion in axion backgrounds and, since it violates parity, leads to birefringence. This can generate a relative phase velocity splitting between left and right polarised light, leading to Faraday-like rotation of linearly polarised light \cite{Carroll:1989vb,Harari:1992ea,DeRocco:2018jwe,Chen:2019fsq}. Modifications to the group velocity were considered in \cite{Mohanty:1993nh} where it was suggested this could lead to time-delays for axion profiles around pulsars. It was also proposed that axion backgrounds could lead to polarisation-dependent bending of light \cite{Plascencia:2017kca}. However a more systematic analysis of photon geodesics revealed this could not happen at $\mathcal{O}(g_{a \gamma \gamma})$ \cite{Blas:2019qqp} for a pure axion background.

In fact, the authors of ref.~\cite{Blas:2019qqp} made a very important and general observation. Whilst there are many references which deal with the propagation of light through an axion background, using a constant term $\partial_\mu a$ in the axion-Maxwell's equations $\partial_\mu F^{\mu \nu} = g_{a \gamma \gamma} \partial_\mu a \tilde{F}^{\mu \nu}$, e.g.~\cite{Carroll:1989vb,Alfaro:2009mr}, piecewise constant within discrete regions \cite{Andrianov:1998wj}) or for which its spatial gradients are neglected \cite{Espriu:2011vj,Plascencia:2017kca,Espriu:2014lma}: in general this is not fully realistic since axion backgrounds are dynamical with space and time dependence. A photon propagating through a localised axion clump probes the full spatial structure of the profile and, in this paper, we show that both spatial and temporal gradients are needed to generate, for example, spectral distortions of photons passing through an axion background which vanishes asymptotically at the points of emission and detection. 

Therefore a proper discussion of dispersion must self-consistently take into account variations of momentum, frequency and position along the ray path. This can be achieved by deriving a system of Hamiltonian optics equations \cite{WeinbergWKB} whose solutions give the photon geodesics and capture all the necessary details about group dispersion, refraction, time delays, and frequency shifts. Furthermore, these equations allow one to study general axion backgrounds without the need to assume any hierarchy between temporal and spatial gradients as would happen for non-relativistic backgrounds. 

The present paper is concerned with examining these key observables within the optics equations, extending the discussion of ref.~\cite{Blas:2019qqp} to higher order effects in $g_{a \gamma \gamma}$ and studying a wider class of observable effects.
First, we point out a distinction between leading and higher order corrections to the photon frequency/momentum. We show that, when plasma is present, one can have asymptotic frequency shifts even for localised axion clouds which vanish at the end points of the photon trajectory. By contrast, in the absence of the plasma, at linear order in $g_{a \gamma \gamma}$, there is no cumulative frequency shift, consistent with \cite{Blas:2019qqp}. Instead, the asymptotic frequency shift in the plasma free case occurs at second order in $g_{a \gamma \gamma}$.

Second, we demonstrate group dispersion at linear order in $g_{a \gamma \gamma}$ for background refractive indices (provided here by plasma) $n_0 \neq 1$. We interpret this in terms of polarisation-dependent time delays occurring at $\mathcal{O}(g_{a \gamma \gamma})$. We also discuss higher order group velocity corrections at $\mathcal{O}(g_{a \gamma \gamma}^2)$ and discuss these in terms of group versus signal velocities, relevant for the speed of information transmission via pulses \cite{Milonni}.
Third, we confirm that there is no asymptotic bending of light at $\mathcal{O}(g_{a \gamma \gamma})$ when the background refractive index $n_0=1$ \cite{Blas:2019qqp}, but show that birefringent refraction \textit{does} happen at $\mathcal{O}(g_{a \gamma \gamma})$ when $n_0\neq 1$. 

The structure of the paper is as follows. In sec.~\ref{sec:Dispersion} we derive the dispersion relations for photons in a general axion background with an additional plasma component. Here we set up the geodesic equations and draw an analogy with Hamiltonian perturbation theory as an interpretation of our perturbative expansion. In secs.~\ref{planar} and \ref{vGSec} we study a simple 1+1 dimensional axion background, which is sufficient to illustrate frequency shifts, group dispersion and time delays. Then in \ref{sec:refraction} we work in 2+1 dimensions, which allows us to demonstrate refraction within a plasma at $\mathcal{O}(g_{a \gamma \gamma})$ and without a plasma at $\mathcal{O}(g_{a \gamma \gamma}^2)$. Finally in sec.~\ref{sec:discussion} we summarise our results and speculate on some applications and proposals for future work. 

\textbf{Physical Motivation.}
The work presented here can be used as a general toolkit, which when appropriately applied and extended, offers the potential to probe the existence of axion backgrounds in a variety of ways. There are of course two particularly pertinent optical features of axion backgrounds which are worth emphasising - firstly, that they violate parity by giving birefringent dispersion and secondly, that the observed shifts in frequency, arrival time and refraction will modulate with a multiple of the period of axion oscillations. 
\begin{enumerate}
    \item \textit{Axion profiles.}~We point out here two interesting types of axion backgrounds to which our results might be applied. One is to postulate axion dark matter, either as a virialised QCD axion background of particles, or as an ultralight scalar whose Compton wavelength is of galactic sizes \cite{Hui:2016ltb}. Both these will lead to birefringent and periodic signals oscillating with frequency $2 \pi/m_a$. Another possibility is to use a superradiant black hole background
    \cite{Arvanitaki:2009fg,Arvanitaki:2014wva,Detweiler:1980uk}, where the axion field values can be especially high. For this case, our discussion of optical properties of \textit{distant} sources, which rely on integrated effects, are particularly relevant.  
    \item \textit{Time-delays.}~Time delays from a galactic/astrophysical axion background could possibly be probed via precision pulsar timing, see e.g. \cite{Khmelnitsky:2013lxt}, although here they arise through a \textit{direct} coupling of the axion field to the photon, rather than via geometric distortions. Equally, a local axion background could be studied in a terrestrial timing experiment,  possibly via interferometry \cite{DeRocco:2018jwe}.
    
     \item \textit{Spectral shifts.}~Our analysis of integrated spectral distortions from axion backgrounds along all or part of the line of sight suggests that, for a line signal from an astrophysical source, one would observe a periodic shift in the position of the the peak frequency.  For a linearly polarised signal, a double image in frequency space for each polarisation would be produced, with the two peaks experiencing periodic oscillation. This could be due -- as with time delays -- to a dark matter background. Another possibility is to observe the spectrum of a superradiant black hole accretion disc whose emission passes through the axion background. In both cases the spectrum would be shifted periodically, with lower frequencies experiencing a greater magnitude of shifting. One could also hope to observe the simpler plasma-free spectral distortions at $\mathcal{O}(g_{a \gamma \gamma})$ in e.g. a terrestrial experiment as suggested in ref.~\cite{Blas:2019qqp} where one measures the frequency changes between emission and detection. 
     
     \textit{Refraction.}~Our results show that birefringent refraction (light-bending) can happen at $\mathcal{O}(g_{a \gamma \gamma})$ when there is a non-trivial background refractive index supplied by, e.g.~plasma. The scenario of polarisation-dependent refraction by a superradiant black hole was considered in \cite{Plascencia:2017kca}, but ruled out at leading-order in $g_{a \gamma \gamma}$ by a more detailed analysis \cite{Blas:2019qqp} of the optics equations. However, in light of our results, it could be interesting to re-run the analysis of ref.~\cite{Plascencia:2017kca} with a plasma component from an accretion disc or interstellar medium and compare the refraction angles to resolutions of upcoming and current telescopes. 
\end{enumerate}

\section{Dispersion in axion backgrounds}\label{sec:Dispersion}
The term \eqref{aGamGam} leads to the following modification of Maxwell's equations %
\begin{align}
\nabla \cdot \textbf{E}& = \rho - g_{a \gamma \gamma} \textbf{B} \cdot \nabla  a\,, \label{Gauss}\\
\nabla \times \textbf{B} - \dot{\textbf{E}} &= \textbf{J} +  g_{a \gamma \gamma}\dot{a} \textbf{B} + g_{a \gamma \gamma} \nabla a \times \textbf{E}\,,\label{curlB}\\
\nabla \cdot \textbf{B} &=0 \label{divB}\,, \\
\dot{\textbf{B} } + \nabla \times \textbf{E}& =0 \label{Bianchi2}\,.
\end{align}
where $\textbf{E}$ and $\textbf{B}$ are the electromagnetic fields and $\textbf{J}$ and $\rho$ are current and charge densities. The electromagnetic fluctuations are related to current fluctuations via 
\begin{equation}
\textbf{J} = \sigma \cdot \textbf{E}, 
\end{equation}
where $\sigma_{ij}$ is a conductivity tensor. Throughout we assume an isotropic and collisionless medium for which the conductivity at a given frequency takes the form $\sigma_{ij}(\omega) = i \delta_{ij} \omega_{\rm p}^2 /\omega$ where $\omega_{\rm p}^2 = n_e e^2 /m_e$ is the plasma frequency-squared and $n_e$ the number density of electrons. This could arise for instance from the interstellar medium, the companion star of a black hole or perhaps an accretion disc, though the following discussion remains general. One could also keep in mind the possibility of any medium which endows the photon with a non-trivial refractive index, which achieves the same end. This may be relevant in exploring extensions of the following analysis to laboratory setups. Some straightforward manipulations of Maxwell's equations we obtain
\begin{align}
&\square \textbf{E} + \nabla(\nabla \cdot \textbf{E}) + \sigma \dot{\textbf{E}} + g_{a \gamma \gamma} \frac{d}{dt} \left[\dot{a}  \textbf{B} + \nabla a \times \textbf{E} \right] =0,\\
&\square \textbf{B} + \sigma \dot{\textbf{B}} - g_{a \gamma \gamma} \nabla \times \left[\dot{a} \textbf{B} + \nabla a \times \textbf{E} \right] =0.
\end{align}
The first equation is derived by taking the time-derivative of eq.~\eqref{curlB} and inserting \eqref{Bianchi2}. The second equation follows by taking time derivative of \eqref{Bianchi2} and using \eqref{curlB} to eliminate $\dot{\textbf{E}}$. Next we perform a Wentzel-Kramers-Brillouin (WKB) expansion, defined by the requirement that the photon wavelength should be less than the gradient scales of the axion background. Formally this amounts to the limit $\partial_\mu \partial_\nu a/\partial_\rho a \ll  \partial_\mu \textbf{E}/\textbf{E}, \partial_\mu \textbf{B}/\textbf{B}$. This requirement allows us to consider solutions of the form
\begin{equation}\label{Sols}
\textbf{E} = \textbf{E}_0 e^{i S}, \qquad \textbf{B}= \textbf{B}_0 e^{i S},
\end{equation}
where frequency and momentum are identified as $\omega = -  \dot{S}$ and $\textbf{k} = \nabla S$. The eikonal approximation is defined by neglecting derivatives $\omega,\textbf{k}$. This is sufficient to obtain dispersion relations and determine rays. With this, we can use plasma conservation $\dot{\rho} + \nabla \cdot \textbf{J} = \dot{\rho} + \sigma \nabla \cdot \textbf{E} =0$ to express
\begin{equation}
\nabla \cdot \textbf{E} = - \left(1- \frac{\omega_{\rm p}^2}{\omega^2} \right)^{-1} g_{a \gamma \gamma} \textbf{B} \cdot \nabla a,
\end{equation}
and use the latter to eliminate the $\nabla (\nabla \cdot \textbf{E})$ term, obtaining:
\begin{align}
&\square \textbf{E} + \omega_{\rm p}^2 \textbf{E}- \frac{(\nabla \textbf{B})\cdot  \nabla a }{1-\omega_{\rm p}^2/\omega^2 } + g_{a \gamma \gamma} \left[\dot{a}  \dot{\textbf{B}}  + \nabla a \times \dot{\textbf{E}} \right]\simeq 0, \label{EWAVE0}\\
&\square \textbf{B} + \omega_{\rm p}^2 \textbf{B}  -g_{a \gamma \gamma}  \left[ \dot{a} \nabla \times \textbf{B}  +  \nabla a (\nabla \cdot \textbf{E}) - (\nabla a  \cdot \nabla) \textbf{E}\right],  \nonumber \\
&\simeq 0 \label{BWAVE0}
\end{align}
where we have neglected second derivatives of the axion field in accordance with the WKB limit described above. Upon using \eqref{Sols}, this system can then be written as \cite{WeinbergWKB,Blas:2019qqp}
\begin{equation}\label{eq:constraintM}
\textbf{M}(\omega , \textbf{k})\cdot (\textbf{E},\textbf{B})^T  = 0,
\end{equation}
where $\textbf{M}$ is a matrix whose structure can be read off from eqs.~\eqref{EWAVE0}, \eqref{BWAVE0}. The condition \eqref{eq:constraintM} is equivalent to demanding that an eigenvalue of $\textbf{M}$ must vanish, which gives the dispersion relation of a particular mode. Explicitly, the eigenvalues of \eqref{EWAVE0}-\eqref{BWAVE0} involving the axion-photon coupling are given by:
\begin{widetext}
\begin{equation}\label{eqDpm}
D^{\pm}= k^2 - \omega_{\rm p}^2 \pm
\frac{1}{[\omega^2 - \omega_{\rm p}^2]^{1/2}}\Bigg[
\omega^2 g_{a \gamma \gamma}^2 \left(   (k\cdot \partial a)^2 -k^2 (\partial_\mu  a)^2\right) +\omega_{\rm p}^2 g_{a \gamma \gamma}^2
\left(
\dot{a}^2 k^2 - 2 \dot{a} \omega (k \cdot \partial a) +(\partial_\mu a)^2\omega^2
\right)
\Bigg]^{1/2}.
\end{equation}
\end{widetext}
%
%
where $k^\mu = (\omega, \textbf{k}) $ and $k^2=k_\mu k^\mu$ and $\partial_\mu a = (\dot{a},\nabla a)$.  Note that the plasma terms are not built solely from Lorentz invariants and therefore violate (local) Lorentz invariance, as expected from the presence of plasma. By contrast, the $\omega_{\rm p} =0$ limit leaves a manifestly Lorentz invariant dispersion relation, owing to the fact the axion-Maxwell's equations $\partial_\mu F^{\mu \nu} = g_{a \gamma \gamma}\partial_{\mu} a \tilde{F}^{\mu \nu}$ are covariant. In the plasma-free case, we recover the same dispersion relation (by setting $D^\pm =0$) of refs.~\cite{Carroll:1989vb, Plascencia:2017kca} which reads
\begin{equation}\label{eq:vacdisp}
\omega_{\rm p} =0: \quad k^2 = \pm g_{a \gamma \gamma}\left[(k\cdot \partial a)^2  - k^2 (\partial_\mu a)^2  \right]^{1/2}. 
\end{equation}
In fact this is the standard form of eigenvalues found for a Chern-Simons-like interaction considered in many works \cite{Andrianov:1998wj,Espriu:2011vj,Alfaro:2009mr} and references therein. Note that, in the present work, we assume a sufficiently weak axion field value so that the photon never suffers a tachyonic instability \cite{Andrianov:1998wj}. However it is worth noting that, for sufficiently low momenta and high density axion fields, this could have interesting physical implications \cite{Boskovic:2018lkj}. We speculate more on these in the discussion. 

In many of these references -- especially for instance the classic reference \cite{Carroll:1989vb} -- the vector $\partial_\mu a$ was treated as a constant background quantity in the spirit of Lorentz and CPT violation. However, recently the authors of ref. \cite{Blas:2019qqp} made a very important observation, namely that when the field $a$ is \textit{dynamical}, the phase space trajectories themselves are altered by spatial and temporal gradients. As a result, it is necessary to self-consistently account the evolution of the frequency, momentum and position of the light ray throughout the trajectory, by solving the full set of Hamiltonian optics equations along the lines of ref. \cite{WeinbergWKB}. To set up the system of equations, one first notes that to enforce dispersion relations along rays, the eigenvalues $D^\pm$ must vanish everywhere along trajectories, which implies
\begin{equation}
\frac{d D^\pm}{d \tau} = \frac{\partial D^\pm}{\partial k^\mu} \frac{d k^\mu}{d \tau} +\frac{\partial D^\pm}{\partial x^\mu} \frac{d x^\mu}{d \tau}  =0, \label{HamiltonEqs}
\end{equation}
where $\tau$ is an arbitrary worldline parameter. We then \textit{define} trajectories according to \cite{WeinbergWKB}
\begin{equation}\label{eq:trajectory}
 \frac{ d x^\mu}{d \tau} \equiv - \frac{\partial D^\pm}{\partial k_\mu},  \qquad \frac{ d k^\mu}{d \tau} \equiv \frac{\partial D^\pm}{\partial x_\mu}, 
\end{equation}
where $k^\mu = (\omega, \textbf{k})$ and $k_\mu = (\omega, -\textbf{k})$. These equations are simply generalisations of Hamilton's equations. One can then eliminate $\tau$ and instead use $t$ as a worldline parameter to arrive at
\begin{align}
\frac{d \textbf{x}}{dt} &= - \frac{\partial D^\pm/\partial \textbf{k}}{\partial D^\pm/\partial \omega}  =\frac{\partial \omega}{\partial \textbf{k}}, \label{dxdt}\\
\frac{d \textbf{k}}{dt}& =  \frac{\partial D^\pm/\partial \textbf{x}}{\partial D^\pm/\partial \omega}  =-\frac{\partial \omega}{\partial \textbf{x}},\label{dkdt}\\
\frac{d \omega}{dt} &= - \frac{\partial D^\pm/\partial t}{\partial D^\pm/\partial \omega} = \frac{\partial \omega}{\partial t}.\label{domegadt}
\end{align}
In fact, one notes that all these equations take the form
\begin{equation}
\frac{d f}{dt} = \left\{ f , \omega \right\} + \partial_t f,\vspace{0.05cm}
\end{equation}
where $\left\{ A, B \right\} = \partial_{\textbf{x}} A \partial_{\textbf{k}} B-\partial_{\textbf{k}} A \partial_{\textbf{x}}  B$  is the Poisson bracket. The interpretation of these equations becomes immediately transparent: they are precisely a system of Hamilton (Jacobi) equations where $\omega=\omega(\textbf{x},\textbf{p} ;t)$ can be seen as a one-particle, time-dependent Hamiltonian yielding the particle energy, and $\textbf{x}$ and $\textbf{p}$ as the standard canonical position and momentum variables. We can think of $\omega_0(\textbf{x},\textbf{k},t) \equiv \left|\textbf{k} \right|$ as the unperturbed Hamiltonian which is then perturbed by the axion interaction. This can then formally be treated using Hamilton-Jacobi perturbation theory by constructing a perturbative expansion about the zeroth order solutions
\begin{align}
\omega_0^2 &= \textbf{k}_0^2+\omega_{\rm p}^2,\\ \textbf{x}_0(t) &= \textbf{v}_g^0 \, t + \textbf{x}_i, \\
\textbf{v}_g^0 &= \textbf{k}_0/\omega_0.
\end{align}
In ref. \cite{Blas:2019qqp}, the authors worked essentially to first order in perturbation keeping only terms $\mathcal{O}(g_{a \gamma \gamma})$ in this expansion. In what follows we discuss higher order corrections which reveal a number of interesting properties.

Since the dispersion relation is, in general, quite complicated and involves finding the roots of a 6th order polynomial, we will analyze some special cases where $\omega = \omega(k)$ takes a simple form and can be solved exactly. One can, however, derive the following perturbative expansion of $\omega(\textbf{k})$. The analytic properties of the dispersion relation are in fact rather interesting. This depends on whether or not the first order perturbation in $g_{a \gamma \gamma}$ vanishes which has additional physical implications
\begin{widetext}
\begin{align}
\omega^\pm(\textbf{k}) & =  \left|\textbf{k}\right| \pm \frac{g_{a\gamma \gamma}}{2} \left[ \hat{\textbf{k}} \cdot \nabla a  + \dot{a} \right] \mp g_{a \gamma \gamma}\dot{a}\frac{\omega_{\rm p}^2}{4\left|\textbf{k}\right|^2} + \frac{g_{a \gamma \gamma}^2}{16 \left|\textbf{k}\right|} \left[\dot{a}^2 + (\hat{\textbf{k}}\cdot \nabla a)^2 - 2|\nabla a|^2 \right]  \nonumber \\
&\pm  \frac{g_{a \gamma \gamma}^3}{16 \left|\textbf{k}\right|^2} \left( \frac{(\partial a)^4}{\dot{a} + (\hat{\textbf{k}}\cdot \nabla a)   } +
\nabla a \cdot \hat{\textbf{k}}  \left[\dot{a}^2 + (\hat{\textbf{k}}\cdot \nabla a)^2 - 2|\nabla a|^2 \right] 
 \right)  + \mathcal{O}(g_{a \gamma \gamma}^4,\,\omega_{\rm p}^2 g_{a \gamma \gamma}^2)    \qquad  \label{vgNormal}\\
 \nonumber
 &\text{unless} \qquad  \dot{a} + \hat{\textbf{k}}\cdot \nabla a \simeq  0,\quad \text{with} \quad \theta_{\textbf{k}} \equiv \arccos{\left(\hat{\textbf{k}}\cdot \nabla a)/|\nabla a|\right)}\neq 0,  
\end{align}
\end{widetext}
%
Here we have effectively performed a double expansion in $\omega_{\rm p}^2$ and $g_{a \gamma \gamma}a$. There are, of course, plasma cross-terms at $g_{a \gamma \gamma}^2$ and higher powers but these do not change any of the physics in a major way, representing only small corrections to existing effects. The special case $\dot{a} + \hat{\textbf{k}}\cdot \nabla a \simeq  0$ is studied in depth in \ref{sec:BiMax}.

Formally, the dispersion relation $\omega = \omega(\textbf{k})$ defines a mass-shell hypersurface in phase space described by $D^\pm =0$ and the perturbative dispersion relations above. For a dynamical background, the photon describes a trajectory on this mass-shell surface $(\textbf{x}(t),\textbf{k}(t), \omega(t))$. However, the dispersion relation does not directly tell us what the on-shell trajectories are and requires a little more manipulation to derive a system of Hamilton's equations, as pointed out in \cite{WeinbergWKB}. By differentiating $\omega$ with respect to $\textbf{k}$, $\textbf{x}$ and $t$,  one is able to obtain the precise details of the integral curves of these trajectories. The point is that $\omega(t)$ and $\textbf{k}(t)$ adjust self-consistently in such a way that $\omega = \omega(\textbf{k}(t))$ holds everywhere along the trajectory, where $\omega (\textbf{k}(t))$ is the on-shell dispersion relation for a particular mode.

\section{Spectral Distortions}\label{planar}
One notable case for which the optics equations are easily tractable and the dispersion relation is simple, is a 1+1 dimensional background with
\begin{equation}
a = a(t,x), \quad \text{and} \quad \textbf{k} \parallel \nabla a ,
\end{equation}
such that the photon momentum $\textbf{k}$ is always parallel to axion gradients $\nabla a$. We can then choose coordinates such that
\begin{equation}
\textbf{k}(t)=(k(t),0,0), \quad \textbf{x}(t)=(x(t),0,0).
\end{equation}
In this case $D_{\pm}$ is
\begin{equation}\label{D1plus1}
D^\pm(t,x) = \omega^2 - k^2 - \omega_{\rm p}^2 \pm g_{a \gamma \gamma}(k \dot{a} + \omega a'),
\end{equation}
where primes denote differentiation with respect to $x$.  The \textit{exact} dispersion relation is then especially simple:
\begin{equation}\label{eq:freq1plus1}
\omega(k) = \left[ k^2 +\omega_{\rm p}^2 \mp g_{a \gamma \gamma}k \dot{a} + \frac{ g_{a \gamma \gamma}^2 a'^2}{2} \right]^{1/2}  \mp \frac{g_{a \gamma \gamma}a'}{2}.
\end{equation}
This case is particularly interesting since it allows us to study the passage of an axion background localised in space. Furthermore, it is sufficient to study group dispersion, time delays and frequency/momentum shifts. 
The geodesic equations then read
\begin{align}
&\frac{dx}{dt} = \frac{2 k \mp g_{a \gamma \gamma}\dot{a}}{2 \omega \pm  g_{a \gamma \gamma}a'},  \\
&\frac{dk}{dt} = \pm g_{a \gamma \gamma} \frac{k \dot{a}' + \omega a''}{2 \omega \pm g_{a \gamma \gamma}a'}, \label{dkdtPlanar} \\
&\frac{d\omega}{dt} = \mp g_{a \gamma \gamma}\frac{k \ddot{a} + \omega \dot{a}'}{2 \omega \pm g_{a \gamma \gamma}a'} . \label{dwdtPlanar}
\end{align}

\subsection{Spectral distortions in external emission}
We now discuss the possibility of detecting spectral shifts when there is \textit{no axion profile} at the end points of the trajectory. By perturbing about a background solution $(\omega_0,k_0)$, from eqs.~\eqref{dkdtPlanar}-\eqref{dwdtPlanar}, we obtain the following expression for the frequency evolution up to $\mathcal{O}(g_{a \gamma \gamma}^2)$
\begin{align}\label{dOmegadtg2}
\frac{d \omega^\pm}{dt} &= \mp\frac{g_{a \gamma \gamma}}{2}\left[n_0 \ddot{a}+  \dot{a}'  \right]  -\frac{g_{a \gamma \gamma}^2\partial_t (\partial  a)^2}{4 k_0} \pm \mathcal{O}(g^3_{a \gamma  \gamma}), 
\end{align}
where
\begin{equation}
n_0 = \frac{k_0}{\omega_0} = \frac{k_0}{[k_0^2 + \omega_{\rm p}^2]^{1/2}} ,
\end{equation}
is the refractive index of the background plasma associated to the trajectory of an unperturbed ray. Note this is also equal to the unperturbed group velocity $v^0_g = d x_0/dt = k_0/\omega_0$ for a plasma. Thus by integration we have that
\begin{equation}
\omega^\pm(t) = \omega_i^\pm + \delta \omega^\pm(t),
\end{equation}
where $\omega^\pm_i$ is the initial frequency and 
\begin{align}
    \delta \omega^\pm (t) =& \pm \Delta \omega_p + \Delta \omega_a\pm \mathcal{O}(g_{a \gamma \gamma}^3),
\end{align}
with
\begin{align}
&\Delta \omega_{p} = - \frac{g_{a \gamma \gamma}}{2}\int^{t}_{0} dt' \left[n_0 \ddot{a}[t',x_0(t')] +  \dot{a}'[t,x_0(t')]  \right], \label{eq:deltaOmega1}  \\
&\Delta \omega_a = -\frac{g_{a \gamma\gamma}^2}{4 k_0} \int^{t}_{0} dt' \partial_t (\partial a)^2, \label{eq:deltaOmega2}
\end{align}
where $x_0(t)$ is the unperturbed photon trajectory - a straight line with constant phase and group velocity $v^0_g = dx_0/dt = k_0/\omega_0 = n_0$,
\begin{equation}
x_0(t) = v_g^0 t + x_{\rm i}.
\end{equation}
We note from the form of eqs.~\eqref{dOmegadtg2}, \eqref{eq:deltaOmega1} and \eqref{eq:deltaOmega2} , when the background has both space and time-dependence, at least one of the terms is not a total derivative. Therefore a localised axion profile which vanishes at the end points of the photon trajectory can impart a spectral shift onto the photon. Indeed one can see this is a general property from eqs.~\eqref{dkdt} and \eqref{domegadt}, such that if \textit{both} space \textit{and} time translation symmetry are broken, the frequency and momentum shifts are prevented from being total derivatives. This is also apparent from simple kinematic arguments by considering a photon with initial/final frequencies and momenta $(\omega_{\rm i},\textbf{k}_{\rm i})$ and $(\omega_{\rm f},\textbf{k}_{\rm f})$, respectively, where the momentum transfer is provided by the axion background. If, asymptotically, the photon satisfies the axion-free mass-shell condition $\omega_{\rm i, f}^2 = (k_{\rm i,f}^2 + \omega_{\rm p}^2)$, then an overall shift in frequency after passing through a local axion region must be accompanied by a shift in momentum, which requires both space and time translation symmetry to be broken. On the other hand, if the background has only one of space or time dependence, it is easy to see that, after a little algebra and using $x_0(t) = t v_g^0 + x_0$, the integrands in \eqref{eq:deltaOmega1} and \eqref{eq:deltaOmega2} can be written as a total derivatives, so $\Delta \omega_p$ and $\Delta \omega_a$ depend only on the end points of the axion trajectory, preventing any  asymptotic frequency shift for a localised axion background. 

We now describe more explicitly at which order in $g_{a \gamma \gamma}$ the asymptotic frequency shift can occur for a localised axion background. In the case where  $n_0 \neq 1$, the leading order asymptotic frequency shift for an asymptotically vanishing axion profile is given by \eqref{eq:deltaOmega1}, which is prevented from being a total derivative by the non-trivial background refractive index. Consequently, the term in \eqref{eq:deltaOmega1} depends on the axion profile along the whole trajectory and not simply the surface terms. This should be compared with the case where the background refractive index is $n_0=1$, as would happen in the absence of plasma,  in which case the leading order frequency shift depends only on the end-points of the trajectory  \cite{Blas:2019qqp}
\begin{align}\label{FreqVac}
&n_0=1: \nonumber \\
& \omega^\pm(t) = \mp \frac{g_{a \gamma \gamma}}{2}\left[ \dot{a}[t,x_0(t)] -\dot{a}[t_i,x_i]\right]   +\omega_i^{\pm} + \mathcal{O}(g_{a \gamma \gamma}^2).
\end{align}
In this case, the asymptotic frequency shift, depending on the whole trajectory history, still occurs but instead at $\mathcal{O}(g_{a \gamma \gamma}^2)$ where it is given by \eqref{eq:deltaOmega2}.

\begin{figure}
    \centering
    \includegraphics[scale=0.6]{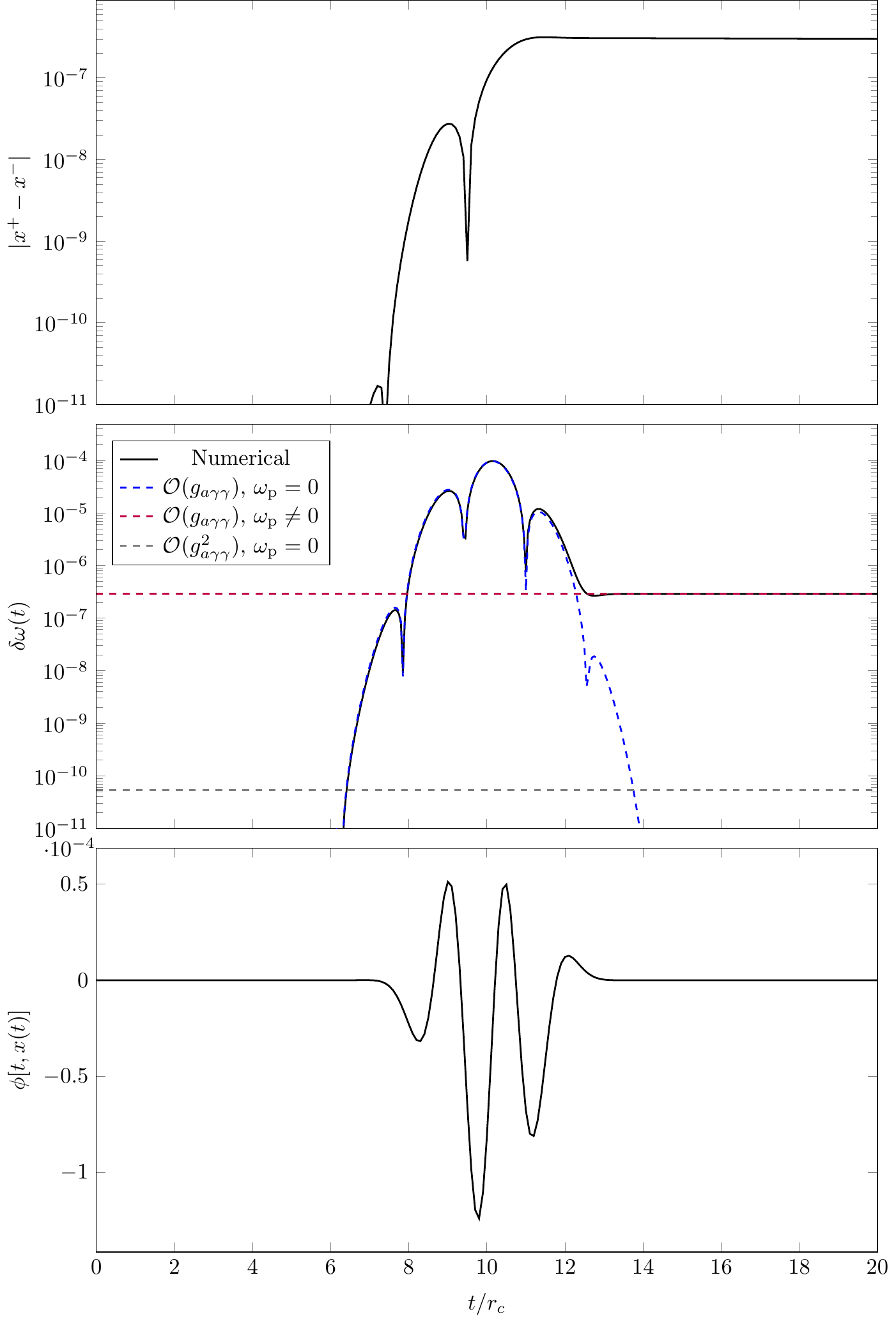}
    \caption{\textit{Top panel}. Relative position/time splitting between left-right polarised photons.  \textit{Middle panel}. Frequency shift for the left-polarised mode. We show the analytic results with and without plasma for comparison. The analytic results in the legend correspond to  eq.~\eqref{FreqVac} (blue-dashed) as well as the asymptotic frequency shifts from eqs.~\eqref{eq:deltaOmega1} and \eqref{eq:deltaOmega2} shown as the dashed horizontal lines.  \textit{Bottom panel}. Axion profile evaluated along trajectory. We used the profile $a = a_0\sin(m_a t) e^{-x^2/r_c^2}$ with $g_{a \gamma \gamma}a_0 = 10^{-4}$, $m_a r_c = 0.5$ and $k_0 r_c=5$. The plasma density is $\omega_{\rm p}/k_0 = 0.08$ and the frequency and position are given in units with $r_c=1$.}
    \label{Traj1by1}
\end{figure}

\begin{figure}
    \centering
    \includegraphics[scale=0.9]{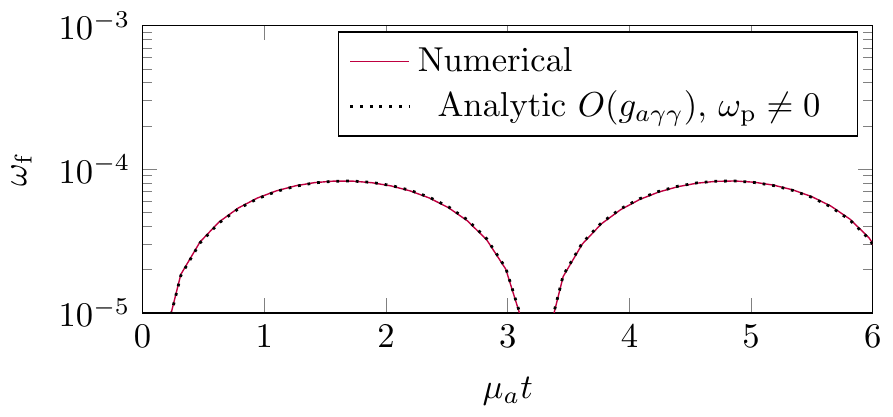}
    \caption{Modulation of final frequency shift $\left|\omega_{\rm f}\right| =\left| \omega(t) - \omega_0\right|$ at the point of detection for passage through a distant axion cloud which vanishes at teh points of emission and detection. Other values are as for fig.~\ref{Traj1by1}. }
    \label{fig:1by1MOdulation}
\end{figure}

This is a very important point. It means that a distant axion profile can impart spectral shifts onto the photon, even if the profile vanishes at the asymptotic points of emission and detection. The different scenarios can be seen from the middle panel of fig.~\ref{Traj1by1}. One especially interesting feature is that the frequency shift as measured at detection will modulate with the oscillation phase of the axion background, as shown in fig.~\ref{fig:1by1MOdulation}.

 It is clearly also worth investigating whether any background which endows the photon with non-trivial refractive index will lead to the same conclusions. This could have particular relevance for dielectric media in laboratory settings.

\subsection{Frequency modulations from local axion backgrounds}

We now consider the possibility in which an ALP field encompasses the point of emission and detection at either end of an experiment. In this case, assuming one can measure the frequency at the point of emission and detection within some apparatus, we need only make a local measurement of $\omega$ so that we can consider the plasma-free case  for which the frequency shift then corresponds to a total derivative, giving only surface terms  
\begin{equation}
     \omega^\pm_{\rm f} - \omega^{\pm}_{\rm i} = \mp \frac{g_{a \gamma \gamma}}{2}\left[ \dot{a}[t_{\rm f},x_{\rm f}] -\dot{a}[t_{\rm i},x_{\rm i}]\right].
\end{equation}
To measure a modulation of the observed frequency, it is then sufficient to have an axion background at \textit{either} the emission or detection points. For the purposes of illustration, consider a homogeneous axion field
\begin{equation}
a(t) = a_0 \sin (m_a t),
\end{equation}
and an apparatus in which the effective distance travelled between emission and detection is $L$, then the change in frequency between emission and detection is
\begin{equation}\label{eq:deltaomega}
       \Delta \omega \equiv \omega^\pm_{\rm f} - \omega^{\pm}_{\rm i} = g_{a \gamma\gamma}\left[\dot{a}(t_{i} + L/c) - \dot{a}(t_{i})\right].
\end{equation}
Clearly this is maximised by tuning $L/c$ to be half the oscillation time, such that \eqref{eq:deltaomega} is the difference between a local minimum and maximum in the oscillation cycle. We then have, estimating the field amplitude via $\rho_{\rm DM} \simeq m_a^2 a_0^2/2$:
\begin{align}
    &\frac{\Delta \omega_{\rm DM}}{\omega} \simeq \nonumber \\
    &1.5 \times 10^{-16}\Bigg[ \frac{g_{a \gamma \gamma}}{10^{-12}\text{GeV}}\Bigg]\Bigg[ \frac{\text{GHz}}{\omega}\Bigg]\Bigg[ \frac{\rho_{\rm DM}}{0.3 \text{GeV}/{\rm cm}^3}\Bigg]^{1/2}.
\end{align}
as reported in ref.~\cite{Blas:2019qqp}. Furthermore we require sufficient temporal resolution $t_{\rm res} \simeq 2\pi/\omega$, to resolve the axion oscillation. This  implies $m_a \lesssim \omega$, which is in any case required by the WKB approximation. This gives an upper bound on the frequency shift
\begin{align}
    &\frac{\Delta \omega_{\rm max,DM}}{\omega}\nonumber \\
    &\lesssim 7 \times  10^{-17}\Bigg[ \frac{g_{a \gamma \gamma}}{10^{-12}\text{ GeV}}\Bigg]\Bigg[ \frac{10^{-5}\text{ eV}}{m_a} \Bigg]\Bigg[ \frac{\rho_{\rm DM}}{0.3 \text{ GeV}/{\rm cm}^3}\Bigg]^{1/2}.
\end{align}
The effective distance traversed by the photon is then of order $L \simeq 2\pi/m_a$ which gives
\begin{equation}
L \simeq 1.25 {\rm m} \Bigg[
\frac{ 10^{-5} \rm{ eV}  }{m_a}\Bigg].
\end{equation}
Presumably in a realistic setup one would use some kind of axion interferometer and study the time-modulation of the interference pattern as in ref.~\cite{DeRocco:2018jwe}. Cavity searches have also been considered \cite{Obata:2018vvr,Liu:2018icu}.

Of course there is a caveat to the above mass range, which is clearly suggestive of QCD axion dark matter. The optics analysis above treats the object $\partial_\mu a$ appearing in the wave equation as a macroscopic coherent classical field background. We remark that ultra-light scalar dark matter has much lower values $m_a \simeq 10^{-22}\text{eV}$ \cite{Hui:2016ltb} which would increase the oscillation times to the scales probed in ref.~\cite{Khmelnitsky:2013lxt}. Meanwhile, a dark matter axion in $m_a \text{eV}$ range would in principle consist of many individual axions. Whilst it has been argued that cold dark matter axions can form an effective Bose condensate \cite{Davidson:2014hfa}, one should still carry out a formal analysis of photon propagation in a background of cold dark matter axions. This would settle the question of  (a) whether a localised cloud of individual axions can be treated effectively as a classical field inserted directly into the photon wave equation leading to  the same analysis above and (b) whether they exhibit coherent oscillations within the detector sizes and timescales. We do not carry out such a treatment in the present work, but merely point out this is something which should be formally verified by, for example, deriving the one-loop photon propagator in a finite density axion background with a given velocity distribution.

\section{Group dispersion, birefringence and time delays}\label{vGSec}
After extracting appropriate powers of $g_{a \gamma \gamma }$ from all quantities involved,  including momentum, we obtain the following expression for the group velocity perturbed about the zeroth order trajectory
\begin{align}
\frac{dx^\pm}{dt}  &=  1 - \frac{\omega_{\rm p}^2}{2 k_0^2} \pm\frac{\omega_{p}^2(a' - \dot{a})g_{a \gamma \gamma}}{2k_0^3} + \frac{g_{a \gamma \gamma}^2 (\partial_\mu a)^2 }{8k_0^2} \nonumber \\
&\pm \frac{g_{a \gamma \gamma}^3(\partial_\mu a)^2\left[  \dot{a} - a'\right]}{8 k_0^3}+ \mathcal{O}(g_{a \gamma \gamma}^4,\omega_{\rm p}^2 g^2_{a \gamma \gamma}), 
\end{align}
 Note therefore that there \textit{is} a birefringent group dispersion at $\mathcal{O}(g_{a \gamma \gamma}$) in the presence of plasma.

This can be seen in fig.~\ref{Traj1by1} numerically with a localised profile which vanishes at $x \rightarrow \pm \infty$. Since the 1D example contains no transverse gradients, photons travel in straight lines but with varying group velocity. Thus  the group arrival times are:
\begin{equation}\label{GroupArrTimes}
t = \int  \frac{dt'}{v_g(t')}.
\end{equation}
Assuming the presence of plasma, we find the following axion contribution to the time-delay 
\begin{align} \label{eq:Deltat}
\Delta t_{p}   &= \mp \frac{g_{a \gamma \gamma}}{4k_0} \frac{\omega_{\rm p}^2}{k_0^2}  \int^{t_{\rm f}}_{0}d t'  \left[a' - \dot{a}\right]. 
\end{align}
This corresponds to the leading-order group velocity
\begin{equation}
v_g =  1 - \frac{\omega_{\rm p}^2}{k_0^2}  \pm \frac{g_{a \gamma \gamma}}{4k_0} \frac{\omega_{\rm p}^2}{k_0^2}    \left[a' - \dot{a}\right] + \mathcal{O}(g_{a \gamma \gamma}^3).
\end{equation}
In this instance, the background plasma ensures that the group velocity is less than the speed of light, provided that the product $g_{a \gamma \gamma} a$ is small, as it must be in order not to violate perturbation theory. The time-delay is similar to that in a magnetised plasma, where the background magnetic field provides the breaking of isotropy,  giving a group velocity \cite{Suresh:2018ayj}
\begin{equation}\label{VgB}
v_g = 1 - \frac{\omega_{p}^2}{k_0^2}   \pm \frac{2\omega_{p}^2 \omega_B}{k_0^3}, \qquad \omega_B = e |\textbf{B}|/m_e.   
\end{equation}
We elucidate further in appendix \ref{sec:BiMax} some more similarities between axion and  magnetic birefringence. We note in passing, with an eye to future astrophysical applications, that the axion birefringent time-delays can be distinguished from the magnetic version via the modulation of the axion signal due to axion oscillations, even though the two effects both run as $\propto 1/k_0^3$. 

Therefore, in the plasma case, $v_g<1$, so it seems reasonable to assume that $v_g$ can be treated as the speed of information transport, thus a meaningful observable for timing experiments.  The system is just a massive vector particle in an anisotropic background. However, one should be cautious with identifying the group velocity with the speed of signal propagation in all instances, as we now discuss.

Note that in the absence of plasma one has $v_g(x,t) \simeq 1 + g_{a \gamma \gamma}^2 (\partial_\mu a)^2/8 k_0^2$. Thus for timelike fields, $(\partial_\mu a)^2 > 0$, we have $v_g>1$, \textit{i.e.} a superluminal group velocity. This is in fact not forbidden  \cite{Milonni,Peters1988}. However, in this case, the group velocity should not be interpreted as the speed at which information propagates \cite{Diener:1996mj,Peters1988}. Instead the \textit{signal speed}, rather than the group or phase velocity should be used for inferring the arrival of information   \cite{Brillouin:1960tos,Milonni,Shore:2003jx,Shore:2003jx}. The signal velocity derivation is more subtle and cannot be inferred in a simple way from the dispersion relation as for the phase and group velocity. Instead, it requires a more detailed treatment of the detection apparatus and the shape of the waveform to be detected \cite{Milonni}. We therefore leave such considerations for future work.

\section{Refraction}\label{sec:refraction}
\begin{figure*}[t!]
    \includegraphics[width = 0.75\textwidth]{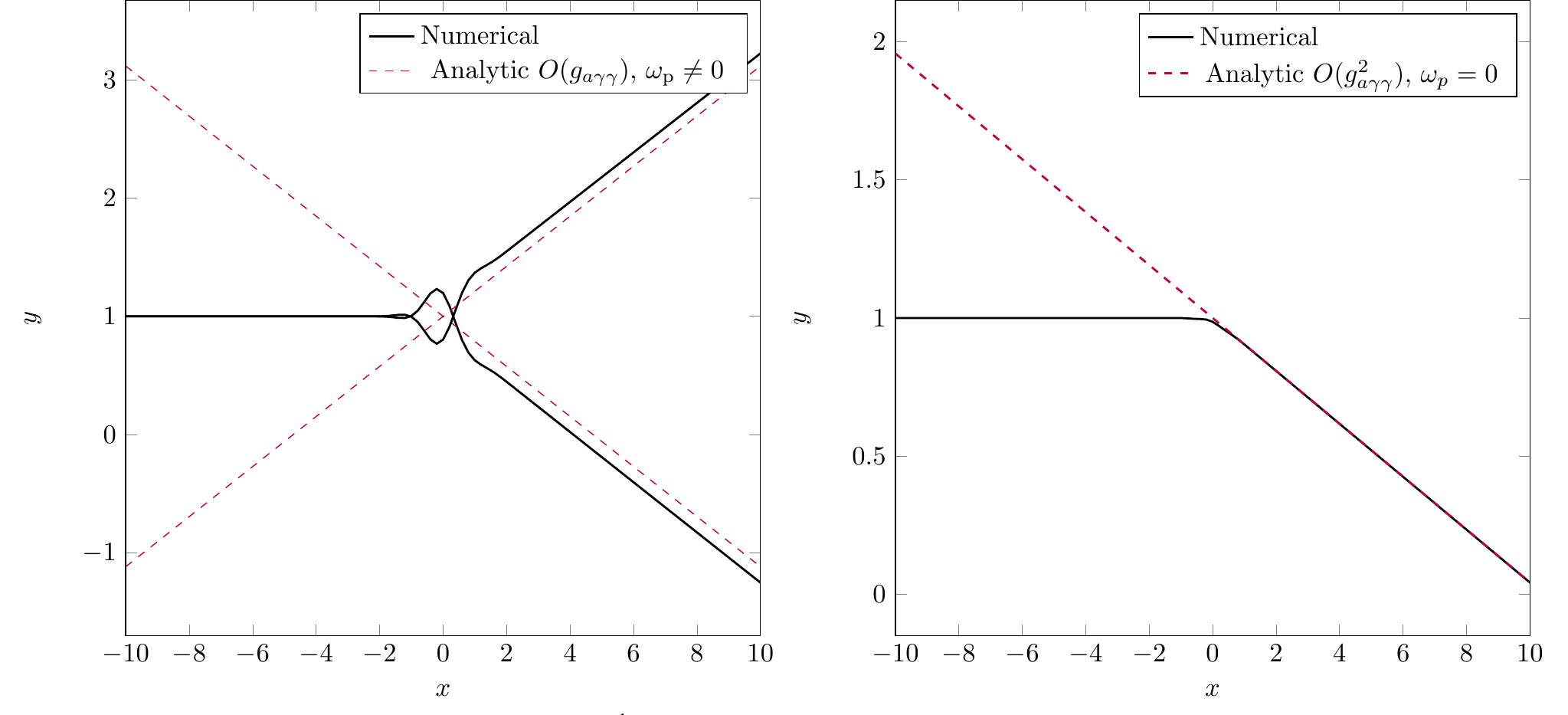}
    \caption{\textit{Left panel.} Refraction through an axion cloud of left/right modes (black lines) with plasma, compared to the asymptotic trajectory (red-dashed) predicted by eq.~\eqref{eq:thetaPL}. The black lines show the rays $(x(t),y(t))$ given by solving the geodesic equations \eqref{dxdt}-\eqref{domegadt} numerically. The axes have been arbitrarily re-scaled for visibility. We chose an impact parameter $b=r_c$ and and a localized Gaussian axion profile, $a = a_0 \sin(m_a t)\exp({-(x^2+y^2)/r_c^2})$, for illustration purposes. With $g_{a \gamma \gamma}a_0 =5 \times 10^{-6}$ and $m_a r_c=0.3$, $k_0 r_c=30$ and a plasma mass $\omega_{\rm p} /|\textbf{k}_0| =0.01$. \textit{Right panel.} Refraction though an axion background without plasma, $\omega_{\rm p}=0$, with comparison to the asymptotic trajectory (red-dashed) predicted by eq.~\eqref{eq:theta}. The other values are the same. Rays propagate from the left to the right of the plot in both cases. }
    \label{plasmabending}
\end{figure*}
We now discuss refraction for passage through a localised axion profile. For this, it is sufficient to consider a 2+1 dimensional background. Normally ray bending is described in terms of the refractive index $n = |\textbf{k}|/\omega$, which is the inverse of the phase velocity $\textbf{v}_p = (\omega/|\textbf{k}|)\hat{\textbf{k}}$. If the photon is to experience deflection after passing through a localised axion profile, it must have differently orientated asymptotic phase velocities.

\subsection{Refraction with plasma}
By using the $\mathcal{O}(g_{a \gamma \gamma})$ solution for $\textbf{k}^\pm$,  we find the leading order corrections to the momentum is
%
\begin{align}
\frac{d\textbf{k}}{dt} & = \pm \frac{g_{a \gamma \gamma}}{2} \left[   n_0 \nabla \dot{a}+ (\hat{\textbf{k}}_0 \cdot \nabla) \nabla a  \right].
\end{align}

When $n_0 \neq 1$, the right-hand side is \textit{not} a total derivative: upon integration, it  is \textit{not} simply given by the surface terms associated with the asymptotic field values at $x_i$ and $x_f$. This allows momentum changes through a localised axion background and therefore refraction to occur at $\mathcal{O}(g_{a \gamma \gamma})$ for non-trivial refractive indices. Integrating the momentum along the trajectory, the transverse component of the outgoing wave vector gives the deflection angle
\begin{equation}\label{eq:thetaPL}
\sin \theta = \pm \frac{g_{a \gamma \gamma}}{2|\textbf{k}_0|} \int_0^t dt'\left[   n_0 \nabla_\perp \dot{a}+ (\hat{\textbf{k}}_0 \cdot \nabla) \nabla_\perp a  \right],
\end{equation}
where $\nabla_\perp$ is the gradient normal to the direction of initial propagation. \eqref{eq:thetaPL} shows that \textit{the existence of a plasma can induce polarisation-dependent ray-bending at leading order} $\mathcal{O}(g_{a \gamma \gamma})$. Indeed this will hold whenever the medium endows the photon with constant refractive index different from one. In the limit $\omega_{\rm p}=0$, one has $n_0=1$ and this integral becomes a total derivative, with the surface terms vanishing for localised axion profile, and so there is no bending in this case. This reproduces the conclusions of ref.~\cite{Blas:2019qqp} that there is no chiral bending at $\mathcal{O}(g_{a \gamma \gamma})$ for a pure-axion background. 

We verified the validity of this approximation by solving the full geodesic equations in 2+1 dimensions. The results of which are shown in fig.~\ref{plasmabending}.

\subsection{Refraction without plasma}
When the refractive index is 1, the momentum and frequency evolution are total derivatives at $\mathcal{O}(g_{a \gamma \gamma})$. Therefore, the non-trivial asymptotic shift in momentum is given at next-to-leading order
\begin{align}
\delta \textbf{k}(t) = - \frac{ g^2_{a \gamma \gamma}}{8|\textbf{k}_0|} \int_0^t dt'  \left( \textbf{k}_0 (\textbf{k}_0 \cdot \nabla) + \hat{\textbf{k}}^{\perp}_0 \nabla_\perp \right)  [ (\partial a)^2].
\end{align}
where $\textbf{k}_0^\perp$ is a unit vector normal to the initial vector. 
%
The angle of deflection is then given by
\begin{equation}\label{eq:theta}
\sin \theta =-\frac{g_{a \gamma \gamma}^2}{8 |\textbf{k}_0|^2} \int_0^\infty dt'\nabla_\perp (\partial a)^2 .
\end{equation}
The analytic and numerical results in this case are compared in fig \ref{plasmabending}.



\section{Discussion}\label{sec:discussion}
In this paper we have examined the dispersive properties of axion backgrounds which lead to spectral distortions, time-delays, and refraction of crossing light. We studied these effects by solving the geodesic equations of Hamiltonian optics equations, deriving both numerical and analytic results. 


Building on the work of Blas \textit{et al.} \cite{Blas:2019qqp}, we verified that i) there is no $\mathcal{O}(g_{a \gamma \gamma})$ asymptotic deflection of light for pure axion backgrounds, but that $\mathcal{O}(g_{a \gamma \gamma})$ birefringent deflection can occur for a non-trivial refractive index of the unperturbed background, induced here by the presence of plasma. ii) In the absence of a non-trivial background refractive index, we found that $\mathcal{O}(g_{a \gamma \gamma}^2)$ refraction does occur. It is unsurprising that refraction should occur at higher order, since there is no symmetry principle to prevent it as the background is both anisotropic and inhomogeneous. 

Perhaps the former effect, induced by the plasma, can have some implications in terms of reviving the idea of ref.~\cite{Plascencia:2017kca} where if a sufficiently dense plasma surrounds the black hole, the refraction generated by a non-trivial axion profile could still be probed at linear order in $g_{a \gamma \gamma}$. It is also worth considering the possibility of probing geodesic deviations from vacuum trajectories in laboratory settings as a means to probe local axion backgrounds. 

iii) A non-trivial constant background refractive index (provided here by a homogeneous plasma) led to asymptotic frequency shifts in the photon, even for localised axion clumps - see \eqref{eq:deltaOmega1}-\eqref{eq:deltaOmega2} and fig. 1. These results have implications both for precision interferometry experiments in a lab, which could in principle probe deviations from an unperturbed photon path and also have relevance for timing-delays and spectral distortions of astrophysical sources. For example, a line signal could exhibit a period spectral shift which oscillates with the axion mass or one could investigate pulsar timing. 

Since plasmas are ubiquitous in an astrophysical setting, we expect the aforementioned results to hold in a large variety of situations. 

Another possibility that we have not considered here is the tachyonic instability for dense axion fields/low frequency photons \cite{Carroll:1989vb,Andrianov:1998wj,Boskovic:2018lkj,Domcke:2019qmm,Domcke:2019mnd} where, for one mode, the frequency becomes imaginary. This can lead in a relative intensity shift between the two modes as one is amplified relative to the other - i.e. dichroism \cite{Marsh:2015xka}.

\textit{Interpretation of plasma/non-lightlike refractive index.} A summary of the order at which the various effects occur is presented in table \ref{table}.
\begin{table}
\begin{tabular}{ |p{3cm}||p{2cm}|p{2cm}|  }
 \hline
 \multicolumn{3}{|c|}{Optical properties of axion-backgrounds} \\
 \hline
 Observable  & no plasma &  plasma \\
 \hline
 $(1-n)$ & $\mathcal{O}(g_{a \gamma \gamma})$ & $\mathcal{O}(\omega_{p}^2)$\\
 $\delta \omega(t\rightarrow \infty)$ & $\mathcal{O}(g_{a \gamma \gamma}^2)$ & $\mathcal{O}(g_{a \gamma \gamma} \omega_{p}^2)$  \\
 $\delta v_g$  &  $\mathcal{O}(g_{a \gamma \gamma}^2)$ &  $\mathcal{O}(g_{a \gamma \gamma} \omega_{p}^2)$\\
 $\delta \theta(t\rightarrow \infty)$  &  $\mathcal{O}(g_{a \gamma \gamma}^2)$   &  $\mathcal{O}(g_{a \gamma \gamma} \omega_{p}^2)$\\
 \hline
\end{tabular}
\caption{Refractive index at a given point, asymptotic frequency shift, group velocity and asymptotic angle of refraction. Note the asymptotic values assume the axion profile vanishes at both emission and detection. The two columns on the right give the order in perturbation at which the effects happen with and without plasma.}
\label{table}
\end{table}
This suggests that to produce an overall asymptotic frequency/momentum shift, time delay or angle of refraction, one requires a refractive index different from $1$. Apparently this can either be provided by the axion at $\mathcal{O}(g_{a \gamma \gamma})$ which is then combined with $\mathcal{O}(g_{a \gamma \gamma})$ coupling to the background, this leads to all asymptotic effects being $\mathcal{O}(g_{a \gamma \gamma}^2)$ and polarisation independent, as these are even powers of $g_{a \gamma \gamma}$. Meanwhile, if there is a non-trivial refractive index provided by the background, it is not necessary to generate a non-trivial refractive index with the axion background, and effects enter at $\mathcal{O}(g_{a \gamma \gamma}(1-n_0))$, leading to asymptotic effects which are linear in $g_{a \gamma \gamma}$. Thus it seems that to have asymptotic (and not just local) effects on these observables for a spatially compact axion background, a non-trivial refractive index is \textit{necessary} and must be provided either by the axion field itself at higher order in $g_{a \gamma \gamma}$,  or by a non-trivial refractive background. It is interesting to speculate on the interpretation of this phenomenon. One observation is that a non-unit refractive index corresponds to a finite amount of proper time passing in the photon's rest frame, allowing the necessary ``time" for these distortions to be imprinted asymptotically. It is also analogous to what happens in the magnetic birefringence of eq.~\eqref{VgB}, which seems to follow the same principle.

\section*{Acknowledgements}
J.I.M acknowledges the support of the Alexander von Humboldt foundation and L.B.V is supported by FCT grant PD/BD/140917/2019. We thank Jo{\~a}o Rosa for stimulating discussions during the early stages of this work and for insightful comments on the manuscript. J.I.M is especially grateful for hospitality from the physics department at the University of Aveiro during his stay.  We also thank Valerie Domcke, Francesca Chadha-Day, Bj{\"o}rn Garbrecht and Graham Shore for useful conversations.


\appendix

\section{Alternative Derivation of Dispersion Relations}
Here we explain how to derive the same dispersion relation in the main text using the formulation of Maxwell's equations used in ref. \cite{Blas:2019qqp}. We show how the dispersion relation they obtain is given by dropping terms $\mathcal{O}(g_{a \gamma \gamma}^2)$ in Maxwell's equations, leading to the dispersion relation of \cite{Harari:1992ea}. Conversely, if these terms are kept, we obtain precisely our dispersion relation.

By using Maxwell's equations, and working in the WKB approximation, one arrives at:
\begin{align}
&\square \textbf{E} + \nabla(\nabla \cdot \textbf{E})  + g \left[\dot{a}  \dot{\textbf{B}}  + \nabla a \times \dot{\textbf{E}} \right]=0, \label{EWAVE}\\
&\square \textbf{B}   -g  \left[ \dot{a} \nabla \times \textbf{B}  +  \nabla a (\nabla \cdot \textbf{E}) - \nabla a  \cdot \nabla \textbf{E}\right]   =0, \label{BWAVE}
\end{align}
where we are neglecting plasma effects ($\sigma = 0$) to facilitate the comparison with \cite{Blas:2019qqp}. The authors of ref.~\cite{Blas:2019qqp} then eliminate the terms $\nabla \cdot \textbf{E}$ and $\dot{\textbf{E}}$ from eq.~\eqref{EWAVE} and $\nabla \times \textbf{B}$ and  $\nabla \cdot \textbf{E}$ from eq.~\eqref{BWAVE} using the Maxwell equations \eqref{Gauss}-\eqref{Bianchi2}
\begin{align}
&\square \textbf{E} + g_{a \gamma \gamma}(\dot{a}\dot{\textbf{B}} - \nabla a \cdot \nabla \textbf{B}) \nonumber \\
&-g_{a \gamma \gamma}^2 \left(\nabla a (\nabla a \cdot \textbf{E}) - (\nabla a)^2 \textbf{E} + \dot{a}\nabla a \times \textbf{B}  \right) =0, \\
&\square \textbf{B} - g_{a \gamma \gamma}(\dot{a}\dot{\textbf{E}} - \nabla a \cdot \nabla \textbf{E}) \nonumber \\
&-g_{a \gamma \gamma}^2 \left(-\nabla a (\nabla a \cdot \textbf{B}) + \dot{a}^2 \textbf{B} + \dot{a}\nabla a \times \textbf{E}  \right) =0,
\end{align}
and then discard terms $\mathcal{O}(g_{a \gamma \gamma}^2(\partial a)^2)$  (similar to \cite{Harari:1992ea}) to obtain (eq.~(8) ref.~ \cite{Blas:2019qqp}):
\begin{align}
&\square \textbf{E} + g_{a \gamma \gamma}(\dot{a}\dot{\textbf{B}} - \nabla a \cdot \nabla \textbf{B}) =0, \label{Etrunc} \\
&\square \textbf{B} - g_{a \gamma \gamma}(\dot{a}\dot{\textbf{E}} - \nabla a \cdot \nabla \textbf{E}) =0. \label{Btrunc}
\end{align}
The authors then use the eigenvalues of the wave operator defined in eqs. \eqref{Etrunc} and \eqref{Btrunc} to obtain the dispersion relation $k^2 = \pm g_{a \gamma \gamma} \partial a \cdot k$, which is missing a term in comparison to our eq.~ \eqref{eq:vacdisp} and refs.~\cite{Carroll:1989vb,Plascencia:2017kca}
\begin{equation}\label{eq:dispfinal}
    k^2 = \pm g_{a \gamma \gamma}\sqrt{(k \cdot \partial a)^2 - k^2(\partial a)^2}.
\end{equation}
Note that we must have $k^2 = \mathcal{O}(g_{a \gamma \gamma})$, as such the second term in the square root is, in total, $\mathcal{O}(g_{a \gamma \gamma}^2)$. Thus we have 
\begin{equation}
    k^2 = \pm g_{a \gamma \gamma} \left( \omega \dot{a} + \textbf{k} \cdot \nabla a\right) + \mathcal{O}(g_{a \gamma \gamma}^2),
\end{equation}
which gives precisely the $\mathcal{O}(g_{a \gamma \gamma})$ form of dispersion relation of ref. \cite{Blas:2019qqp,Harari:1992ea}. One can use this dispersion relation as long as one never goes beyond linear order in $g_{a \gamma \gamma}$. However, since are also interested in higher order effect, we keep all terms and use the general form eq.~\eqref{eq:dispfinal}. 
 
We also point out that there are two perturbative expansions in this paper
\begin{equation}\label{eq:PT}
g_{a \gamma \gamma} (\partial a)/k_\gamma \sim (g_{a \gamma \gamma}a) (m_a/k_\gamma) \ll 1,
\end{equation}
where $m_a$ is the mass of the axion field and
\begin{equation}\label{eq:WKB}
 \partial_\mu \partial_\nu a/\partial_\rho a \ll  \partial_\mu \textbf{E}/\textbf{E}, \partial_\mu \textbf{B}/\textbf{B} \implies k_a \ll k_\gamma,
\end{equation}
which should not be confused. The first describes the validity of perturbation theory and controls the convergence of an expansion counted by the powers of the dimensionless parameter $g_{a\gamma \gamma} a$. This is precisely the Hamilton-Jacobi perturbation series. Meanwhile, the second limit corresponds to a WKB approximation in which the photon frequency is larger than the axion frequency. Note that for large field amplitudes and axion couplings, $g_{a \gamma \gamma} a > 1$, but still small axion frequencies, it is formally possible to violate \eqref{eq:PT} but obey \eqref{eq:WKB}.

\section{Other modes} \label{sec:BiMax}
For
\begin{equation}
\hat{\textbf{k}}\cdot \nabla a + \dot{a} = 0, \quad \text{and} \quad \theta_{\textbf{k}} \neq 0,
\end{equation}
there are two possible dispersion relations with each left/right mode choosing the same branch:
\begin{align}
  \text{O}:\quad &\omega_O  = |\textbf{k}|,\label{O} \\
  \text{X}: \quad &\omega_X = |\textbf{k}| + g_{a \gamma \gamma}^2\frac{ \sin^2 \theta_\textbf{k} |\nabla a|^2}{2\left|\textbf{k}\right|} + \mathcal{O}(g_{a \gamma \gamma}^4, \omega_p^2). \label{X} \qquad 
\end{align}
In this case, since the dispersion is the same for each circularly polarised mode, the phase velocities are identical in each case and the eigenmode corresponds to \textit{linear polarisation}. The direction of the linear polarisation is defined with respect to the background anisotropy. These two linearly polarised modes above are defined as O (ordinary) and X (extraordinary): the X-mode depends on the orientation relative to the background anisotropy described by $\nabla a$. The O-mode propagates as though in absence of the anisotropy, whilst the X-mode does not. 

\begin{figure}[h!]
    \centering
    \includegraphics[scale=0.8]{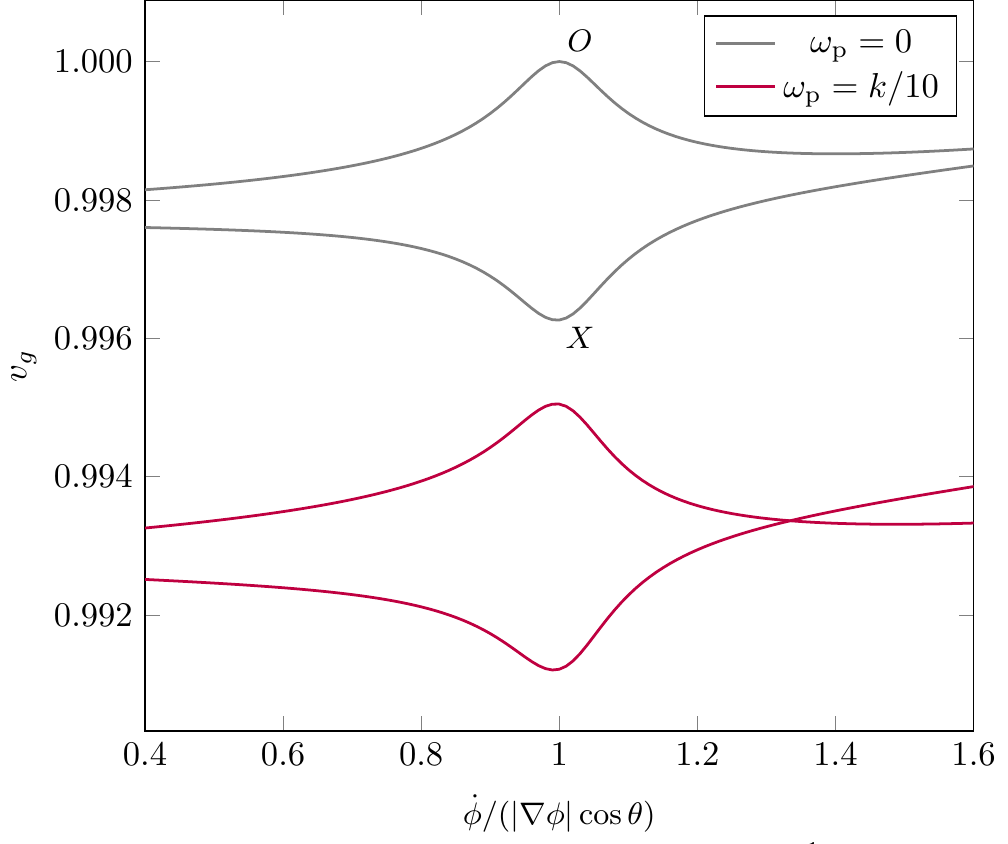}
    \caption{The local group velocities near the O and X modes as a function of the axion field configuration given by numerical solution of  the dispersion relation associated to \eqref{eqDpm}. We took $\theta_{\textbf{k}} =\pi/3$ and $g_{a \gamma \gamma}\dot{a}/k_0 = 10^{-1}$. The group velocity splitting becomes maximal as the photon approaches the extraordinary and ordinary (axion-free) mode, respectively. The top and bottom branches of each pair show the O and X modes respectively.}
    \label{fig:MaximalBiref}
\end{figure}

We verified that \eqref{vgNormal} holds away from the bump in fig.~\ref{fig:MaximalBiref}, at which point the values at each bump are given by \eqref{O} and \eqref{X} modes. This is no different to what happens in a magnetised plasma,  which admits O and X modes defined by their polarisation relative to the background anisotropy defined by the magnetic background field. 
However, we emphasise that the O and X modes are a generic feature of any anisotropic medium: in the present case, the anisotropy is provided by spatial axion gradients rather than a background magnetic field.
%

Note that when $\theta_\textbf{k} =0$ (modulo $\pi$) \textit{and} $\dot{a} + |\nabla a| =0$,  one finds both dispersion relations of the corresponding O and X modes are the trivial vacuum dispersion, as is expected from the limiting case of eq.~\eqref{X}. This can be seen from the dispersion relation (neglecting $\omega_{p}$ terms which do not change the conclusions) $k^2 = \partial a\cdot k$, which gives precisely the condition $(k^\mu \pm \partial^\mu a)^2 =0$ and implies $\omega = |\textbf{k}|$ \cite{Andrianov:1998wj}. However, for a mode which has some perpendicular momentum component relative to $\nabla a$, there can be X modes.

Of course for a dynamical axion field, the O-X mode dispersion relation can only hold instantaneously, since any momentum component normal to the gradients will be deflected. Similarly, for non-relativistic field configurations, the condition can only be fulfilled as the oscillation curve of $\dot{a}$ briefly crosses $\nabla a$.  For most of the trajectory the dispersion relation \eqref{vgNormal} holds and can be used to derive analytic results.

We now note the following enhancement of birefringent effects which happens in critical regions of the axion background which can be described by drawing an analogy with  birefringence in magnetised plasmas which fall under the heading of \textit{magneto-optic effects}. In the main text, we derived the dispersion relations for the case $\dot{a} \neq - \hat{\textbf{k}}\cdot \nabla a$. For the point $\dot{a} = - \hat{\textbf{k}}\cdot \nabla a$ described previously in this Appendix, one finds the following perturbative form of the group velocities, treating $ \partial_\mu a$ as a constant:
\begin{align}\label{vOX}
&v_g^O= 1 - \frac{\omega_{\rm p}^2}{2|\textbf{k}|^2},  \\
&v_g^X=1 -
\frac{ \omega_{\rm p}^2 }{2 |\textbf{k}|^2}
- 
\frac{g_{a \gamma \gamma}^2 \left|\nabla a\right|^2}{2|\textbf{k}|^2} 
\sin^2 \theta_{\textbf{k}},
\end{align}
Note the first of these resembles dispersion in the absence of an axion background (O-mode), whist the second experiences an anisotropy. These are generic modes of anisotropic media and we now describe them in more detail. 

\subsection{O and X modes in magnetised plasma}
Suppose a dilute, weakly magnetised plasma for which the conductivity takes the form
\begin{equation}
 \sigma_\textbf{B} \cdot \textbf{E} = \frac{\omega_{\rm p}^2}{\omega} \textbf{E} + i \textbf{g}_\textbf{B} \times \textbf{E},
\end{equation}
where $\textbf{g}_\textbf{B} = (\omega_{\rm p}^2/\omega^2) (e \textbf{B}/m_e)/\omega$ is the \textit{gyration vector}, which points in the direction of $\textbf{B}$. The corresponding wave equation is then
\begin{equation}\label{gyration}
(\omega^2 - k^2 - \omega_{\rm p}^2) \textbf{E} + \textbf{k}(\textbf{k} \cdot \textbf{E}) + i \textbf{g}_{_\textbf{B}} \times \textbf{E}  =0. 
\end{equation}
This admits two modes with which propagate perpendicular to $\textbf{g}_\textbf{B}$ with $\textbf{k}\cdot \textbf{g}_\textbf{B} =0$. These are the O (ordinary) and X (extraordinary) modes. The first gets its name from the fact that its dispersion is identical to that of an unmagnetised plasma since it is polarised with $\textbf{E} \parallel \textbf{g}_B$ in addition to being transverse with $\textbf{k} \cdot \textbf{E}=0 $, $\omega^2 = |\textbf{k}|^2 + \omega^2_{\rm p}$.  Meanwhile, the extraordinary mode is both longitudinally and transversely polarised, propagating perpendicular to $\textbf{g}_\textbf{B}$. Its dispersion relation is $k^2 = (1 - \omega_{\rm p}^2/\omega^2)(1 - \omega_B^2[\omega^2 - \omega_{\rm p}^2 - \omega_B^2]^{-1})$ and thus receives $\mathcal{O}(g_B^2)$ corrections relative to the O-mode. 

This is a general property of dispersion in anisotropic media with a single direction of isotropy breaking. In general, waves propagating normal to the direction of the anisotropy fall into two categories: those parallelly and perpendicularly polarised with respect to the anisotropy. The first of which propagates as though the anisotropy were not there. These two modes typically exhibit the greatest birefringence.

\subsection{O and X modes in axion backgrounds}
Note that, by manipulating Maxwell's equations, we can derive the WKB equation of ref.~\cite{Carroll:1989vb} written purely in terms of $\textbf{E}$
\begin{equation}
(\omega^2 - k^2 - \omega_{\rm p}^2) \textbf{E} + \textbf{k}(\textbf{k} \cdot \textbf{E}) + i \textbf{g}_a \times \textbf{E}  =0, 
\end{equation}
where $\textbf{g}_a$, the effective \textit{axion gyration vector}, is
\begin{equation}
\textbf{g}_a = g_{a \gamma \gamma}\left[\dot{a}\, \textbf{k} +  \omega \nabla a  \right].
\end{equation}
Thus in regions where $\textbf{k}\cdot \textbf{g}_a =0$, we obtain the corresponding O and X modes. Specifically, for a photon trajectory with a given $\textbf{k}$, this occurs when
\begin{equation}
\textbf{g}_{ {\rm eff } }  \cdot \hat{\textbf{k}} =0 \implies  \left( |\textbf{k}| \dot{a}  + \omega  |\nabla a| \cos \theta_\textbf{k} \right)=0. \label{eq:resReg}
\end{equation}
%

Let us now discuss under what conditions \eqref{eq:resReg} can actually be satisfied. Consider first an on-shell axion plane wave $a \sim \cos(\omega t - \textbf{k} \cdot \textbf{x})$ and $ \omega = m_a^2 + |\textbf{k}|^2$. Such a propagating solution is ``timelike" in the sense that $|\dot{a}| > |\nabla a| $ is always strictly true for $m_a \neq 0$. 

However, the axion configuration around a superradiant black hole \cite{Detweiler:1980uk} is a bound state solution and therefore allows for spacelike $\partial_\mu a$, leading to the existence of regions in which $\dot a \simeq -(\omega/|\textbf{k}|)\textbf{k} \cdot \nabla  a =0$. It is also important to point out that, although the amplitude of fluctuations of $\dot{a} \sim \omega a$ is greater than $\nabla a$, since both terms are sinusoidal, there remain regions where the two terms can cancel exactly.  However, the time for which this condition holds is typically short-lived, especially for a non-relativistic axion background where for most of the trajectory $\dot{a} \gg \nabla a$.
\bibliography{main}{}

\end{document}